\documentclass[runningheads]{llncs}
\usepackage{graphicx}
\usepackage{amsmath}
\usepackage{amssymb}
\usepackage{bm}
\usepackage{mathrsfs}
\usepackage{color,xcolor}
\usepackage{booktabs}
\usepackage{tabularx}
\usepackage{multirow}
\usepackage{hyperref}
\usepackage{marvosym}
\hypersetup{colorlinks=true,
            linkcolor=blue,
            anchorcolor=blue,
            citecolor=blue}
\begin{document}
\title{A Point Cloud Generative Model via Tree-Structured Graph Convolutions for 3D Brain Shape Reconstruction}
\titlerunning{Brain Shape Reconstruction Based on Tree-Structured Graph Convolutions}
%
%
%
%
%
%

\author{Bowen Hu\inst{1,2} \and
Baiying Lei\inst{3} \and
Yanyan Shen\inst{1} \and
Yong Liu\inst{4} \and
Shuqiang Wang\inst{1}(\textrm{\Letter})}

\institute{1 Shenzhen Institutes of Advanced Technology, Chinese Academy of Sciences,
Shenzhen, Guangdong, China\\
\email{sq.wang@siat.ac.cn}\\
2 University of Chinese Academy of Sciences, Beijing, China\\
3 Shenzhen University, Shenzhen, Guangdong, China\\
4 Renmin University of China, Beijing, China\\
}

\maketitle

\begin{abstract}

Fusing medical images and the corresponding 3D shape representation can provide complementary information and microstructure details to improve the operational performance and accuracy in brain surgery. However, compared to the substantial image data, it is almost impossible to obtain the intraoperative 3D shape information by using physical methods such as sensor scanning, especially in minimally invasive surgery and robot-guided surgery. In this paper, a general generative adversarial network (GAN) architecture based on graph convolutional networks is proposed to reconstruct the 3D point clouds (PCs) of brains by using one single 2D image, thus relieving the limitation of acquiring 3D shape data during surgery. Specifically, a tree-structured generative mechanism is constructed to use the latent vector effectively and transfer features between hidden layers accurately. With the proposed generative model, a spontaneous image-to-PC conversion is finished in real-time. Competitive qualitative and quantitative experimental results have been achieved on our model. In multiple evaluation methods, the proposed model outperforms another common point cloud generative model PointOutNet.

\end{abstract}
\section{Introduction}
\label{sec:intro}

With the continuous advancement of medical operation methods, minimally-invasive and robot-guided intervention technology have gradually been used in brain surgery in recent years, bringing patients smaller surgical wounds, shorter recovery time, and better treatment experience. However, the development also brings new requirements. Since doctors cannot directly observe lesions and surgical targets during operations, their experience is often not so efficient. Therefore, these technologies will not be fully mature before the surgical environment and the real-time information acquisition ability improve. Recently, the application of intraoperative MRI (iMRI) has become more and more extensive, and some work used it to relieve the stricter visual restrictions in minimally invasive surgery \cite{1,2}. But unlike the rich internal details of the brain, MRI cannot provide intuitive and visually acceptable information of the surface and the shape of target brains that is more important for surgery. Thus, it is an inevitable direction of development for these types of surgery to find some indirect 3D shape information acquisition methods that are accurate and controllable. Considering the limitations of the use of conventional scanners and the inconvenience of the acquisition of images in brain surgery, these methods should be based on algorithms, rather than physical medical equipment, and can reconstruct the 3D shape of the target from as little traditional information as possible.

There are several alternative representations in the field of 3D surface reconstruction, such as voxels \cite{3}, meshes \cite{4}, and point clouds\cite{29,30}. A point cloud, which is represented as a set of points in 3D space, just uses N vertices to describe the overall shape of the target. As for comparisons, the voxel representation requires cubic space-complexity to describe the reconstruction result. And meshes, each of which will be treated as a domain to registration in the reconstruction, use a $N\times N$ dimensional adjacency matrix. Therefore, it is a reasonable proposal to choose point clouds as the reconstruction representation in surgical scenes that require less time consuming and flexibility.

In previous work, There are many kinds of research on the generative method of point cloud representation. In the PC-to-PC generation field, \cite{5} proposed an auto-encoder (AE) model to fold and recover the point clouds. \cite{6} proposed a variational auto-encoder (VAE) model to generate close point cloud results and then apply them to fracture detection and classification. \cite{7,8,9} proposed different GAN architectures to learn the mapping from Gaussian distribution to multiple classes of point cloud representations, so as to reconstruct point clouds in an unsupervised manner. In the image-to-PC generation field, \cite{10} proposed a new loss function named geometric adversarial loss to reconstruct point clouds representation that better fits the overall shape of images. In \cite{11}, a deep neural network model composed of an encoder and a predictor is proposed. This model named PointOutNet predicts a 3D point cloud shape from a single RGB image. On the basis of this work, \cite{12} applied this model to the one-stage shape instantiation to reconstruct the right ventricle point cloud from a single 2D MRI image, thereby simplifying the two-stage method proposed by \cite{13}.

However, to the best of our knowledge, no effort has been devoted to the development of point cloud reconstruction of brains. Since Deep learning technology has been popularized in the medical prediction\cite{34,35,36}, and has been applied in many fields such as maturity recognition \cite{14,15}, disease analysis \cite{16,17,18,40,41}, data generation \cite{19,20}, there are many works that combine deep learning with 3D data for accurate reconstruction \cite{21,22,23}. Generative adversarial network, as well as many of its variants \cite{24,25,26}, is a widely used generative model and is known for its good generation quality. Using it for image-point cloud conversion can preserve features of images to the greatest extent and improve the accuracy of point clouds. Variational method is another generative model\cite{37,38,39}, and can extract features to provide more reliable guidance for the generation of GAN. Meanwhile, graph convolutional networks (GCN) has achieved great success in solving problems based on graph structures \cite{31,32,33}, and has been proven to be very effective in the generation and analysis of point clouds \cite{7,8}. To solve the problem that it is difficult for the rGAN model in \cite{9} of generating realistic shapes with diversity, we consider construct novel GCN modules to replace the fully connected layer network in rGAN. In this paper, a new GAN architecture is proposed to convert a single 2D brain MRI image to the corresponding point cloud representation. Our model consists of an encoder, a novel generator that is composed of alternate GCN blocks and branching blocks, and a discriminator similar to WGAN-GP in \cite{27}. A tree-structured graph convolutional generative mechanism is constructed to ensure that the generated point cloud guided by image features is as close to the target brain as possible.

The main contribution of this paper is to propose a general and novel GAN architecture to reconstruct accurate point clouds with high computational efficiency. This model is the first work to introduce point cloud generation into the fields of brain image analysis and brain shape reconstruction. We use the variational encoder method and graph convolutional algorithm to build and improve our GAN architect, and use different loss functions to ensure that all parts of the model achieve the best training effect.

\begin{figure*}[t]
\centering
\includegraphics[height=5.2cm, width=12cm]{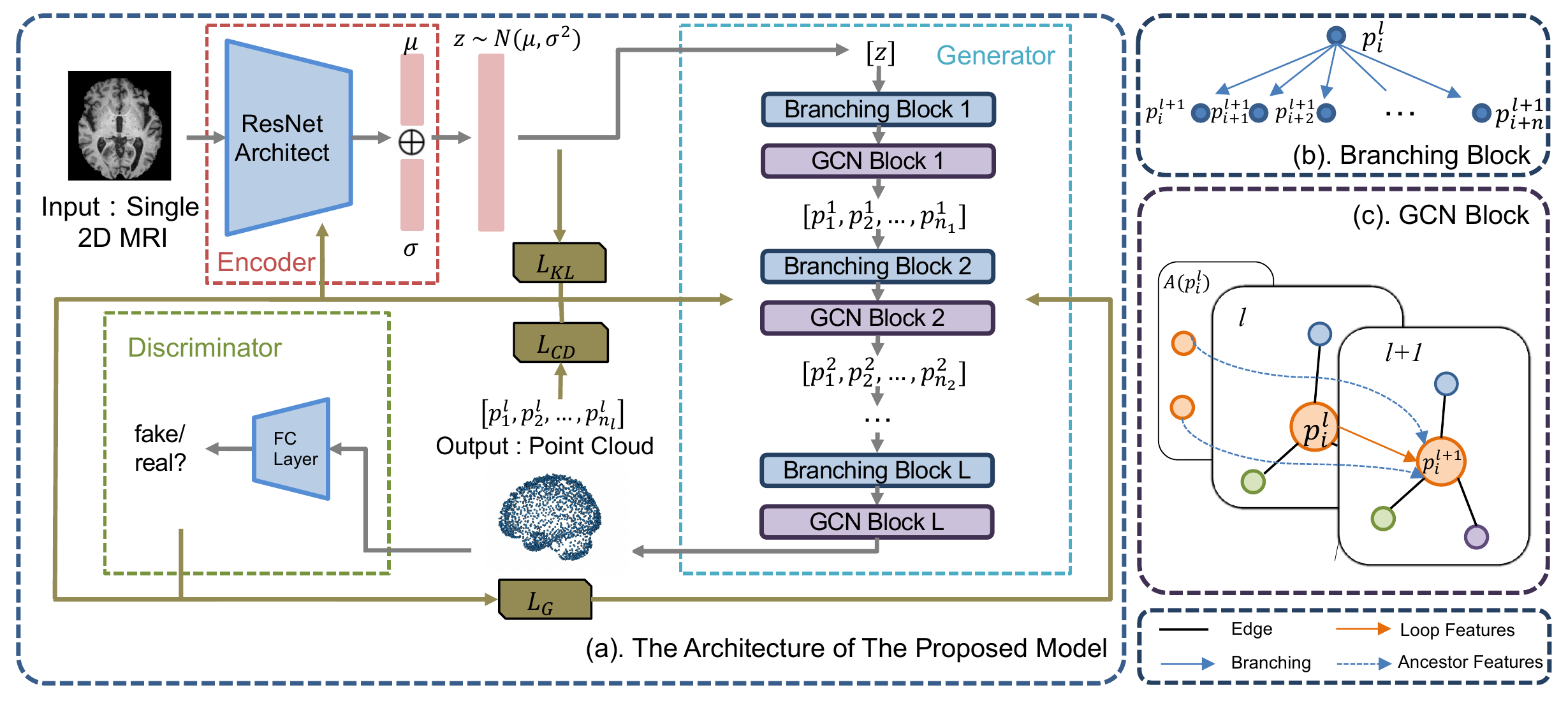}
\caption{The Architecture of The Proposed Model: (a) Pipeline of our model. Gray arrows represent how the information flows transfer, and brown arrows represent the backward propagation direction of the loss function. (b) How branching blocks in generator work. (c) Details of GCN blocks in generator.}
\label{fig:1}
\end{figure*}

\section{Methods}
\label{sec:methods}

\subsection{Generative Architecture Based on Tree-structured GCN}
\label{ssec:GAN in GCN}
Given a specific brain subject, the point cloud of it can be expressed as a matrix $Y_{N\times3}$ which denotes a set of N points and each row vector represents the 3D coordinate of a vertex. Preprocessed 2D images are required for the proposed generative model, which contains three networks, namely encoder, generator, and discriminator. The architecture of the model is shown in Fig. \ref{fig:1}(a). Gray arrows represent how the information flows transfer, and brown arrows represent the backward propagation direction of the loss function. The encoder designed to have a similar architecture to ResNet in \cite{28} takes such images $I_{H\times W}$ as inputs, where H is the height and W is the width of an image. It produces vectors $z \in \mathbb{R}^{96}$ from a Gaussian distribution with a specific mean $\mu$ and standard deviation $\sigma$ as outputs, which are treated as a point set with only one single point by the generator. The generator uses a tree-structured graph network which has a series of GCN blocks and branching blocks to expand and adjust the initial point set. Then discriminator differentiates the output $Y_{N\times3}$ (in this work, $N = 2048$) and real point cloud, and then enhances the generator to make generative point clouds closer to the ground truth. Specially, we use the model in the Wasserstein GAN with the gradient penalty method at the discriminator.

\subsubsection{GCN Block.}
\label{sssec:gcn block}
Some work has used graph convolutional networks for point cloud generation tasks. Inspired by \cite{7,8}, multi-layer improved GCNs are considered in our model to implement an efficient generator. Each layer consists of a GCN block and a branching block. The graph convolutions in GCN blocks (Fig. \ref{fig:1}(c)) is defined as
\begin{equation}
p^{l+1}_i = \sigma \left (\bm{F}^{l}_{K}(p^{l}_i) + \sum\limits _{q_j \in A(p^{l}_i)}U^l_jq_j + b^l\right ),
\end{equation}
where there are three main components: loop term $S^{l+1}_i$, ancestor term $A^{l+1}_i$ and bias $b^l$. $ \sigma (\cdot)$ is the activation function.

Loop term, whose expression is
\begin{equation}
S^{l+1}_i = \bm{F}^{l}_{K}(p^{l}_i),
\end{equation}
is designed to transfer the features of points to the next layer. Instead of using a single parameter matrix $W$ in conventional graph convolutional networks, the loop term uses a K-support fully connected layer $\bm{F}^{l}_{K}$ to represent a more accurate distribution. $\bm{F}^{l}_{K}$ has K nodes $(p^l_{i, 1}, p^l_{i, 2}, ..., p^l_{i, k})$, and can ensure the fitting similarity in the big graph.

Ancestor term allows features to be propagated from the ancestors of a vertex to the corresponding next connected vertex. This term of a graph node in conventional GCN is usually named neighbors term and uses the information of its neighbors, rather than its ancestors. But in this work, point clouds are generated dynamically from a single vector, so the connectivity of the computational graph is unknown. Therefore, this item is modified to
\begin{equation}
A^{l+1}_i = \sum\limits _{q_j \in A(p^{l}_i)}U^l_jq_j,
\end{equation}
to ensure structural information is inherited and multiple types of point clouds can be generated. $A(p^{l}_i)$ is the set of ancestors of a specific point $p^{l}_i$. these ancestors map features spaces from different layers to $p^{l}_i$ by using linear mapping matrix $U^l_j$ and aggregate information to $p^{l}_i$.

\subsubsection{Branching Block.}
\label{sssec:branching block}
Branching is an upsampling process, mapping a single point to more points. Different branching degrees $(d_1, d_2, ..., d_n)$ are used in different branching blocks (Fig. \ref{fig:1}(b)). Given a point $p^{l}_i$, the result of the branching is $d_l$ points. Therefore, after branching, the size of the point set becomes $d_l$ times the upper layer. By controlling the degrees, we ensured that the reconstruction output in this work is accurate 2048 points.

The purpose of branching blocks is to build the tree structure of our generative model, thereby giving the generator the ability to transfer the feature from the root $z \in \mathbb{R}^{96}$ to points at the last layer and generate complex point cloud shapes, which can ensure the accuracy and effectiveness in brain shape reconstruction.

\subsection{Training of the Proposed GAN Model}
\label{ssec:point cloud gan}
In addition to the generator, there are two networks, encoder and discriminator. Fig.\,1 shows the process that the encoder extracts the random vector $z$ from the 2D MRI $I$ and the generator generates the point cloud which was judged by discriminator. Because of the particular data format, conventional loss cannot train this generative network well. Chamfer distance is often used as loss function in point cloud generation. Given two point clouds $Y$ and $Y'$, Chamfer distance is defined as
\begin{equation}
\mathcal{L}_{CD} = \sum\limits_{y' \in Y'} min_{y \in Y} ||y' - y||^2_2 + \sum\limits_{y \in Y} min_{y' \in Y'} ||y - y'||^2_2,
\end{equation}
where $y$ and $y'$ are points in $Y$ and $Y'$, respectively. Thus, we define the loss function of encoder and generator as
\begin{equation}
\mathcal{L}_{E-G} = \lambda_1 \mathcal{L}_{KL} + \lambda_2 \mathcal{L}_{CD} - \mathbb{E}_{z\sim \mathcal{Z}}[D(G(z))],
\end{equation}
where $ \mathcal{L}_{KL}$ is the Kullback-Leibler divergence, $\lambda_1$ and  $\lambda_2$ are variable parameters and $\mathcal{Z}$ is a Gaussian distribution calculated by encoder.

Meanwhile, the loss which is similar with \cite{27} is used on the discriminator. $\mathcal{L}_{D}$ is defined as
\begin{equation}
\mathcal{L}_{D} = \mathbb{E}_{z\sim \mathcal{Z}}[D(G(z))] - \mathbb{E}_{Y\sim \mathcal{R}}[D(Y))] + \lambda_{gp}\mathbb{E}_{\hat{x}}[(||\nabla_{\hat{x}}D(\hat{x})||_2 - 1)^2],
\end{equation}
where $\hat{x}$ are sampled from line segments between real and fake point clouds, $\mathcal{R}$ represents the real point cloud distribution and $\lambda_{gp}$ is a weighting parameter.

\section{Experiments}
\label{sec:experiments}
\subsubsection{Experimental Setting.}
The proposed model is trained on our in-house dataset with voxel-level segmentation converted into point clouds. The preprocessing was prepared under the professional guidance of doctors. The dataset consists of 317 brain MRIs with AD and 723 healthy brain MRIs. All bone structures of MRIs are removed and the remaining images are registered into format $91\times 109\times 91$ by software named FSL. 900 MRIs are randomly selected to construct the training set and the others are used for the test set. We choose some 2D slices of MRIs which were normalized to [0, 1] as the input of the model. 2000 epochs were trained for our generative model and the comparative model, PointOutNet. We use a CPU of Intel Core i9-7960X CPU @ 2.80GHz$\times$32 and a GPU of Nvidia GeForce RTX 2080 Ti for our experiments. We set $\lambda_1$ and $\lambda_{gp}$ to 0.1 and 10, respectively, and use Adam optimisers with an initial learning rate of $1 \times 10^{-4}$. Specific to the generator, we adjust training parameters dynamically by increasing $\lambda_2$ from 0.1 to 1.

\begin{figure}[h]
\includegraphics[width=12cm]{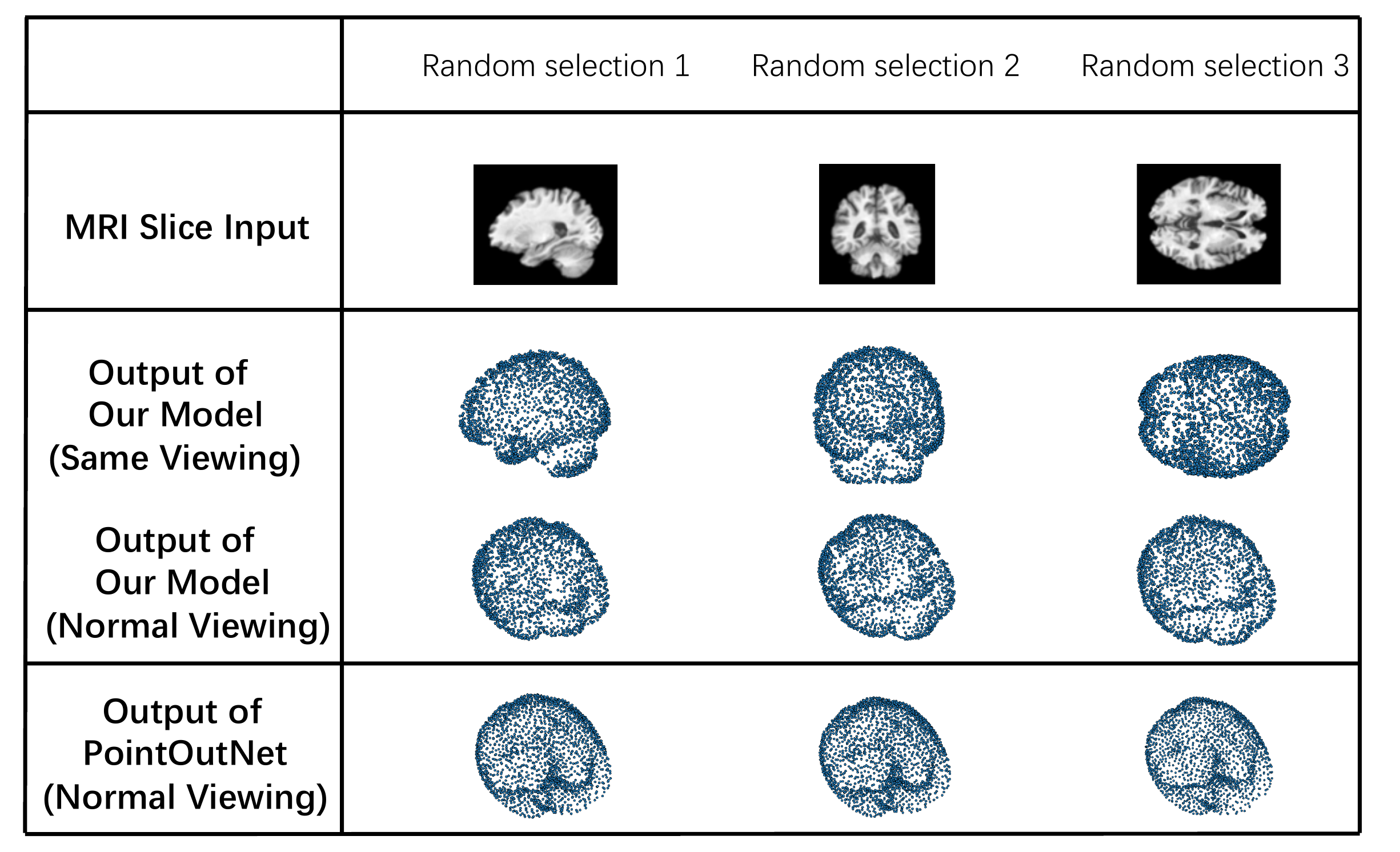}
\caption{Intuitive illustrations of Some Angles of Outputs with Three Views MRI Inputs}
\label{fig:2}
\end{figure}

\begin{figure}[h]

%
\begin{minipage}[b]{1.0\linewidth}
  \centering
  \centerline{\includegraphics[width=8.5cm]{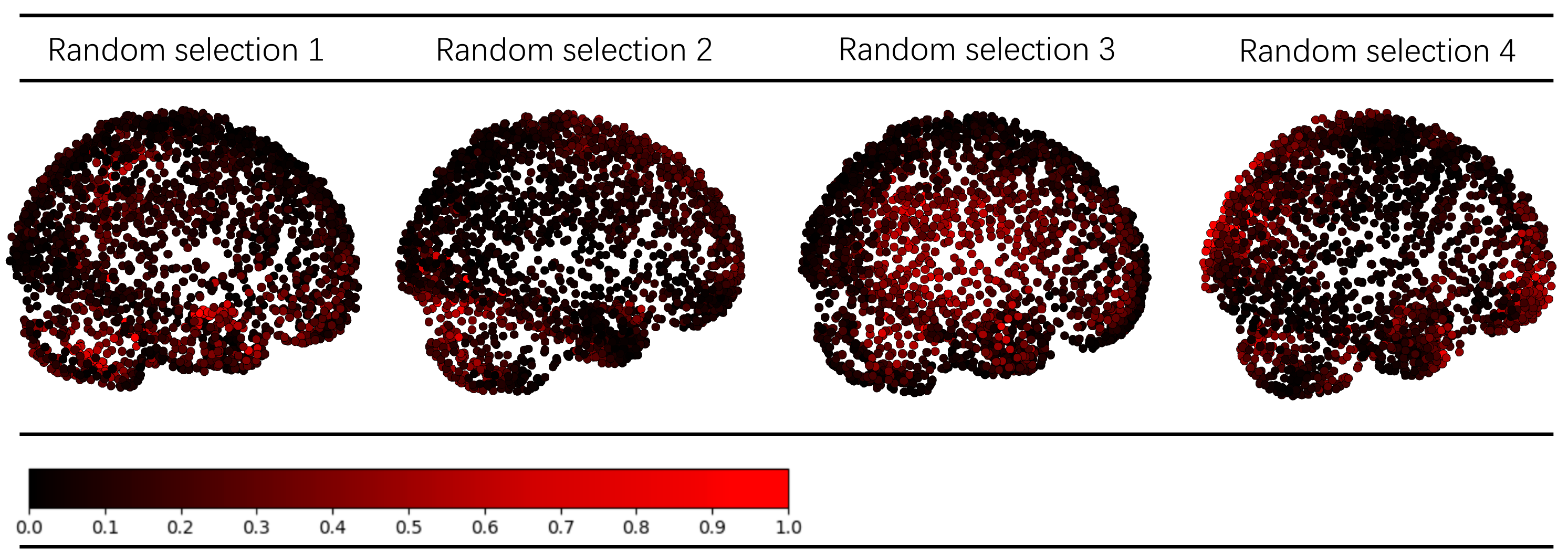}}
  \centerline{(a)}\medskip
\end{minipage}
\begin{minipage}[b]{1.0\linewidth}
  \centering
  \centerline{\includegraphics[width=8.5cm]{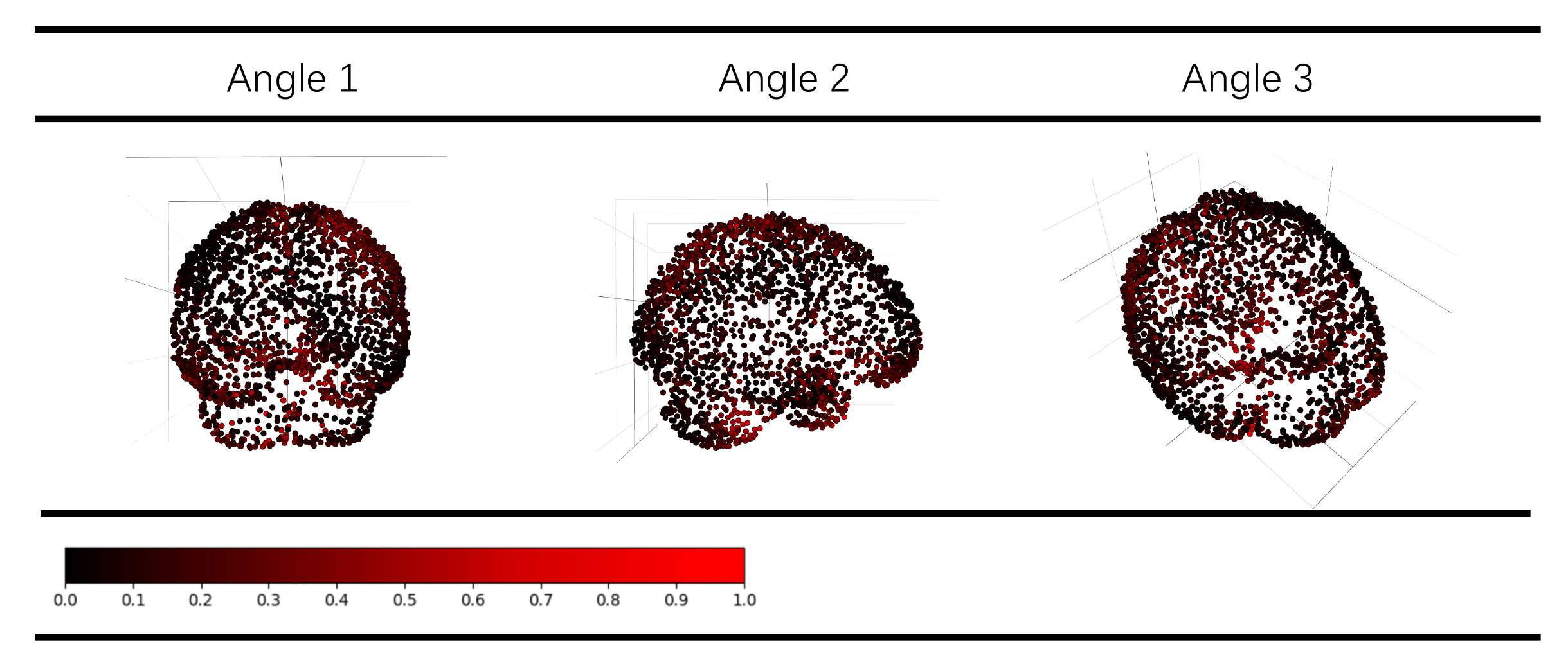}}
  \centerline{(b)}\medskip
\end{minipage}

\caption{Intuitive illustrations of (a) The Generated Point Clouds Whose Vertices is Colored by PC-to-PC Error and (b) A Generated Point Cloud Whose Vertices is Colored by PC-to-PC Error in different angles}
\label{fig:3}
\end{figure}
\subsubsection{Qualitative Evaluation.}
In order to investigate the performance of the proposed model, we conduct some qualitative evaluations and the results are in Fig. \ref{fig:2}, which shows three views about the pairs of MRI inputs and point cloud outputs. In addition, we compared the output of our model with the output of PointOutNet under normal viewing. The point cloud outputs derived from PointOutNet have a highly similar appearance between each other, especially for the description of the overall structure of the brain. The structural features are disappearing and only some details are different. On the contrary, our model has the ability to aggregate point features and generate high-quality and detailed 3D point clouds from slices in multiple directions. Moreover, the outputs of PointOutNet are observed to be unevenly distributed in the 3D space, which is also avoided in our model.

Also, we show the point-by-point accuracy of our generated point cloud in Fig. \ref{fig:3}. Fig. \ref{fig:3}(a) shows the reconstruction effect of different randomly selected objects under the same angle, while Fig. \ref{fig:3}(b) shows the error analysis of the same object under different angles. Each vertex is colored by an index named $PC-to-PC error_{\hat{y}}$ and introduced in \cite{12}. Given two point clouds $Y$ and $Y'$, PC-to-PC error is defined as

\begin{equation}
PC-to-PC Error_{\hat{y}} = min_{y \in Y} ||y' - y||^2_2
\end{equation}

\begin{equation}
PC-to-PC Error_{\hat{Y}} = \sum\limits_{y' \in Y'} min_{y \in Y} ||y' - y||^2_2
\end{equation}

where $y$ and $y'$ are points in $Y$ and $Y'$, respectively. The value on the axis is multiplied by a factor $10^{-4}$. We can see that most of the vertices in the reconstruction results have a very small error.

\subsubsection{Quantitative Measures.}
\begin{figure}[ht]
\includegraphics[width=12cm]{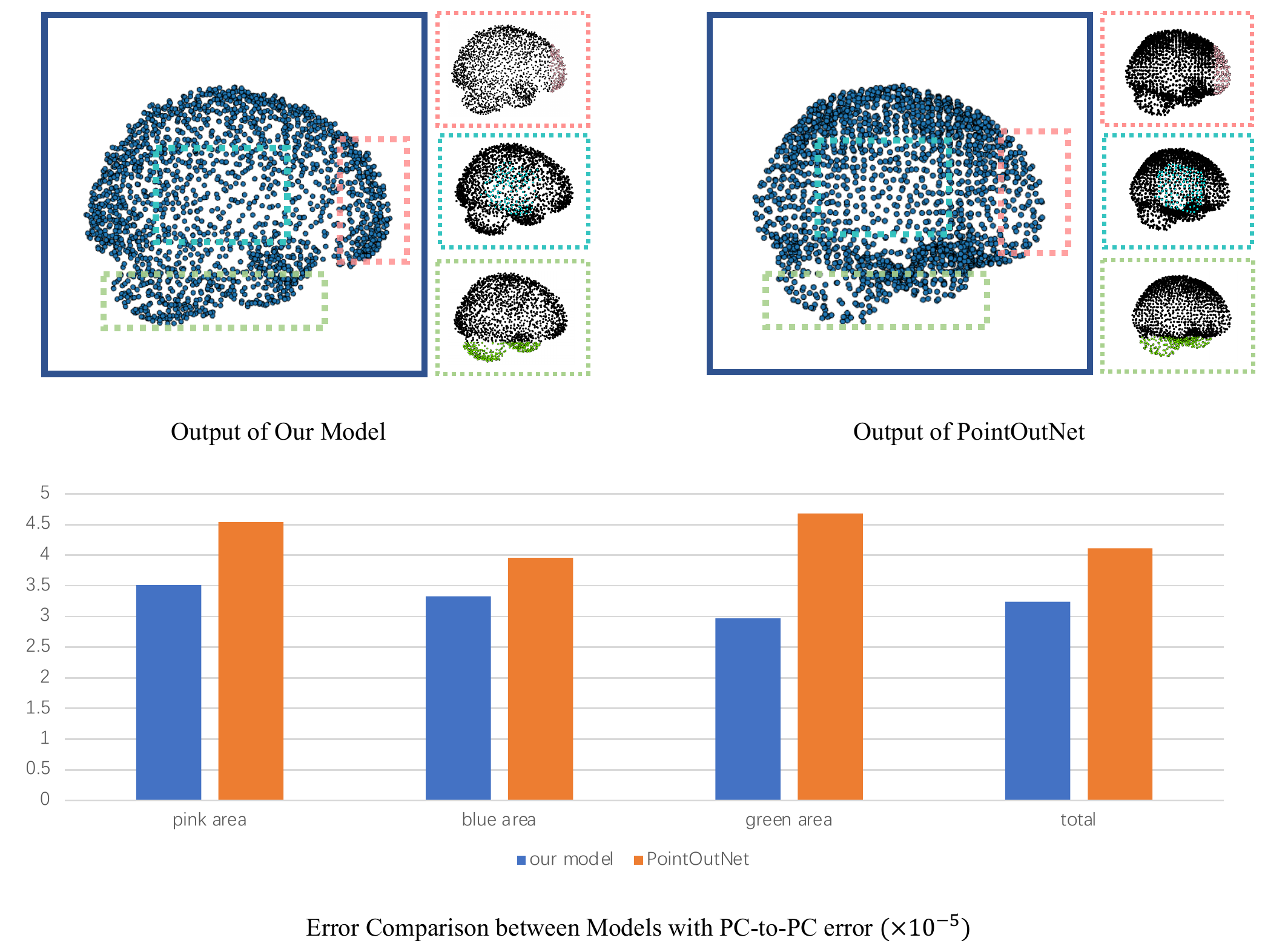}
\caption{The PC-to-PC Error of the Different Area and the Total Point Cloud (Lower is Better).}
\label{fig:4}
\end{figure}

PC-to-PC error and Chamfer distance which are based on correlation and the positional relationship between vertices are employed to quantitatively measure the distance between different point clouds. Also, we apply the earth mover's distance (EMD) introduced in \cite{9} to measure the overall distribution gap between the results of each model and the ground truth. The EMD is defined as
\begin{equation}
\mathcal{D}_{EMD} = min_{\phi: Y \rightarrow Y'} \sum\limits_{x \in Y}||x - \phi(x)||_2
\end{equation}
where $\phi$ is a bijection. In order to comprehensively compare the reconstruction results between our model and PointOutNet, we fairly selected 3 reconstruction key areas and marked them with pink, blue, and green. The relevant results are shown in Fig. \ref{fig:4}. Considering that the number of generated points in each area may be different, we use the PC-to-PC error which is calculated for each point in this evaluation.
\begin{table}[h]
	\centering
	\caption{The More Specifically PC-to-PC Error of the gernerated Point Cloud in Fig. 4.}
	\label{table:1}
	\begin{tabular}{p{2cm}|cccc}
		\toprule
		Area&\color{pink}Pink Area&\color{cyan}Blue Area&\color{green}Green Area&Total \\
		\midrule
		Our Model&\textbf{3.51}&\textbf{3.33}&\textbf{2.97}&\textbf{3.24}\\
		PointOutNet&4.54&3.95&4.68&4.11\\
		\bottomrule
	\end{tabular}
\end{table}
\begin{table}[h]
	\centering
	\caption{Quantitative measures for test. In both metrics, the lower the better.}
	\label{table:2}
	\begin{tabular}{p{2cm}|ccc}
		\toprule
		Metric&Our Model.&Our Model without D.&PointOutNet \\
		\midrule
		CD&\color{red}9.605&\color{blue}13.825&12.472\\
		EMD($\times 10^{-1}$)&\color{red}7.805&11.577&\color{blue}14.329\\
		\bottomrule
	\end{tabular}
\end{table}
This indicator reveals that our method has better performance in the reconstruction of all the important areas and the overall point cloud. The more specific values of the related area are in Table \ref{table:1}. We also calculated the chamfer distance and earth mover's distance of our model and PointOutNet throughout the test. Specifically, we set up an additional model which removes the discriminator and uses Chamfer distance as a loss function to train the encoder and generator for a comparative experiment. We expect to compare this model with the proposed model to prove the advantages of the adversarial reconstruction of our model. Table \ref{table:2} shows results quantitative results of all models using CD and EMD. As can be seen, our model outperforms the compared models with the lowest CD (9.605) and EMD (0.7805), indicating that the accuracy of our model is higher than that of PointOutNet in the test.



\section{Conclusion}
\label{sec:conclusion}
A shape reconstruction model with the GAN architect and the graph convolutional network is proposed in this paper to relieve the visual restrictions in minimally invasive surgery and robot-guided intervention. 3D point cloud is used as the representation. A novel tree-structured generative mechanism is constructed to transfer features between hidden layers accurately and generate point clouds reliably. To the best of our knowledge, this is the first work that achieves the 3D brain point cloud reconstruction model which uses brain medical images and reconstructs the brain shape. The proposed model has a very competitive performance compared to another generative model. In future work, this method will continue to be developed to solve the problem of visual limitation in minimally invasive surgery and other advanced medical techniques.

\section{Acknowledgment}
This work was supported by the National Natural Science Foundations of China under Grant 61872351, the International Science and Technology Cooperation Projects of Guangdong under Grant 2019A050510030, the Distinguished Young Scholars Fund of Guangdong under Grant 2021B1515020019, the Excellent Young Scholars of Shenzhen under Grant RCYX20200714114641211 and Shenzhen Key Basic Research Project under Grant JCYJ20200109115641762.

\end{document}